\documentclass[%
 reprint,
superscriptaddress,
 amsmath,amssymb,
 aps,
]{revtex4-1}

\usepackage{booktabs}
\usepackage{multirow}

\usepackage{subcaption}

\usepackage{bm} 
\usepackage{graphicx}
\usepackage{dcolumn}
\usepackage{bm}
\usepackage{hyperref}

\begin{document}

\preprint{APS/123-QED}

\title{Multiscale probing of a Hernquist-type environmental black hole spacetime with the Sgr A* shadow and S2 orbital dynamics}
\author{Lai Zhao}
\email{laizhao.phys@outlook.com}

\author{Meirong Tang}
\email{tangmr@gzu.edu.cn}

\author{Zheng-Wen Long}
\email{zwlong@gzu.edu.cn (Corresponding author)}

\author{Zhaoyi Xu}%
\email{zyxu@gzu.edu.cn (Corresponding author)}
\affiliation{%
 College of Physics,Guizhou University,Guiyang,550025,China
}%


\begin{abstract}

The supermassive black hole Sgr A* at the Galactic center provides a unique opportunity to probe the distribution of environmental matter around black holes.
In this work, we adopt the Hernquist-type environmental black hole spacetime, a non-vacuum exact solution of the Einstein field equations, as its gravitational model to describe the joint gravitational field of the black hole and its surrounding matter, with environmental effects characterized by the dimensionless compactness $C$ and the characteristic scale $\alpha$.
We combine black hole shadow data with two sets of S2 star data provided by Do et al. and Gillessen et al., and constrain the model parameters using the Markov chain Monte Carlo method. At the 95\% credible upper limit, the shadow-only data constrain $C < 1.498\times10^{-1}$.but provide no effective constraint on $\alpha$. The two S2 datasets yield $C<5.239\times10^{-5}$ and $C<1.303\times10^{-4}$, respectively, with $\alpha$ exhibiting a bimodal structure in both cases. After combining the shadow and S2 star data, the $C$ upper limits are tightened to $C<3.760\times10^{-5}$ and $C<1.073\times10^{-4}$, respectively.
These results indicate that current observations rule out highly compact configurations of the environmental halo, while the obtained constraints are consistent with the typical compactness range of matter halos. However, $\alpha$ still exhibits a significant bimodal degeneracy, indicating that current observations are insufficient to uniquely determine the radial distribution of the environmental halo. Future observations of multiple stellar orbits may provide further insights into the radial structure of the environmental halo.

\begin{description}
\item[Keywords]
Hernquist-type environmental black hole spacetime; Markov chain Monte Carlo analysis; environmental effects
\end{description}
\end{abstract}

\maketitle


\section{\label{sec:level1}Introduction}

The supermassive black hole Sgr A* at the Galactic center provides an ideal laboratory for testing general relativity and investigating the gravitational effects of surrounding matter around black holes. 
The Event Horizon Telescope (EHT) observations of Sgr A* have revealed its black hole shadow on near-horizon scales \cite{EventHorizonTelescope:2022wkp,EventHorizonTelescope:2022apq}, 
while the GRAVITY Collaboration, through long-term astrometric and radial velocity observations of the S2 star orbiting Sgr A*, has probed orbital dynamics at scales of more than one thousand Schwarzschild radii from the pericenter, providing tests of gravitational redshift and Schwarzschild orbital precession \cite{GRAVITY:2018ofz,GRAVITY:2020gka}. 
These two types of observations probe different regions of the strong-field spacetime and provide complementary constraints on the geometry of the same black hole spacetime across multiple scales, which can be used to test gravitational theories and constrain possible spacetime modifications caused by environmental effects.

In existing studies, vacuum black hole spacetimes, such as the Kerr and Schwarzschild solutions, are commonly adopted as benchmark models for Sgr A*, and have been used to investigate observable signatures including black hole shadows, stellar orbital precession, and gravitational redshift \cite{Vagnozzi:2022moj,EventHorizonTelescope:2022wok,EventHorizonTelescope:2022xqj,GRAVITY:2018ofz,GRAVITY:2020gka}. 
Based on these vacuum solutions, previous works have utilized EHT observations and S2 star orbital data to constrain the Kerr or Schwarzschild black hole hypotheses and their possible deviations \cite{EventHorizonTelescope:2022xqj,Navarrete:2026zyu,Vagnozzi:2022moj}. 
Meanwhile, various black hole spacetimes inspired by modified gravity or quantum gravity scenarios have also been extensively investigated \cite{Vagnozzi:2022moj,Afrin:2022ztr,Gogoi:2024vcx,Kalita:2023xlu,Yao:2026pjz,QiQi:2026pnb,Xamidov:2026dps}. 
However, the central regions of galaxies are neither vacuum nor isolated systems, but rather complex environments composed of various matter components, including stars, gas, and dark matter. 
The distribution of these matter components can be described collectively by the energy-momentum tensor, which acts as the source term in Einstein's field equations and determines the spacetime geometry \cite{Genzel:2010zy,Bland-Hawthorn:2016lwg,Sadeghian:2013laa,Figueiredo:2023gas,Fernandes:2025osu,Gondolo:1999ef,Shakura1973}. 
Such environmental effects should not be regarded merely as external perturbations imposed on a vacuum black hole background, but should instead be incorporated into the non-vacuum spacetime geometry. 
With the advent of high-precision observations, such as the next-generation Event Horizon Telescope (ngEHT) and future space-based gravitational-wave detectors \cite{Johnson:2023ynn,Ayzenberg:2023hfw,LISA:2017pwj,TianQin:2015yph,TianQin:2020hid,Hu:2017mde}, these environmental effects may leave imprints on horizon-scale imaging features and stellar orbital dynamics \cite{Barausse:2014tra,Cole:2022yzw,Kouniatalis:2025itj,EventHorizonTelescope:2022wkp,Ghez:2008ms}. 
Therefore, incorporating environmental effects into a unified theoretical and observational modeling framework is essential for precision tests in the strong-field regime, and high-precision observations of Sgr A* provide a unique opportunity to constrain the spacetime structure and the properties of its surrounding matter environment.

In this work, we adopt a black hole spacetime embedded in a Hernquist-type galactic halo as the background model. 
This spacetime was obtained by Cardoso et al. by extending the Einstein cluster method \cite{Einstein:1939ms,Geralico:2012jt} to a Hernquist density profile \cite{Hernquist:1990be}, with an anisotropic fluid serving as the matter source, and solving Einstein's field equations exactly. 
It provides an analytical black hole spacetime solution incorporating environmental effects \cite{Cardoso:2021wlq}, which can describe the combined gravitational field of a central black hole and the surrounding Hernquist-type matter distribution in galactic centers.
Unlike the vacuum Schwarzschild spacetime, this spacetime geometry is determined not only by the central black hole mass $M_{\rm BH}$, but also by the environmental matter mass $M$ and its characteristic distribution scale $a_0$ \cite{Cardoso:2021wlq}. 
The solution preserves the event horizon structure in the near-central region, while on larger scales its gravitational behavior is dominated by the halo matter satisfying the Hernquist density profile, thereby reflecting the modification of spacetime geometry induced by environmental matter. 
Therefore, this model provides a unified analytical spacetime framework that describes both the black hole-dominated inner region and the galactic environmental region, offering a rigorous theoretical basis for multi-scale observational constraints.
Compared with many models that characterize the environment around black holes through Newtonian potential corrections or purely phenomenological parameterizations (see, e.g., Refs. \cite{Xu:2018wow,Konoplya:2019sns,Macedo:2013qea}), the model proposed by Cardoso et al. \cite{Cardoso:2021wlq} consistently describes the gravitational coupling between black holes and surrounding galactic matter within the framework of general relativity. 
A recent comparative analysis by Fauzi et al. \cite{Fauzi:2025yse} further demonstrated the necessity of such a self-consistent framework in the strong-field regime.
This model has been widely applied to theoretical studies of black hole shadows and gravitational lensing \cite{Macedo:2024qky,Xavier:2023exm,Kouniatalis:2025itj}, quasinormal modes \cite{Feng:2025iao,Konoplya:2021ube,Lutfuoglu:2025kqp}, tidal effects \cite{Liu:2022lrg}, orbital dynamics \cite{Tan:2024hzw,Haroon:2025rzx}, quasi-periodic oscillations \cite{Stuchlik:2021gwg}, and gravitational-wave signals \cite{Destounis:2022obl,Dai:2023cft,Rahman:2023sof,Zhang:2024ugv}. 
These studies establish a rigorous theoretical foundation for quantitatively understanding the influence of surrounding matter on black hole shadows and stellar orbital dynamics.

Previous constraints on Sgr A* have mostly relied on a single observational channel. For example, black hole shadow observations or measurements of the orbital precession of the S2 star have been used independently to constrain additional parameters introduced by different black hole models (see, e.g., Refs. \cite{Bambhaniya:2026slo,Navarrete:2026zyu,Sharipov:2025xws,QiQi:2026pnb,Yao:2026pjz,Zeng:2025kqw}). In such analyses, some studies fix the mass $M$, distance $D$, semimajor axis $a$, and eccentricity $e$ to their externally determined best-fit values in order to highlight the sensitivity of the shadow or orbital precession to the additional parameters (see, e.g., Refs. \cite{Sharipov:2025xws,Xamidov:2025prl,Wei:2025qlh,Meng:2025vva,Huang:2024oli,QiQi:2024dwc,Zeng:2025wlb,Zeng:2025kqw}). Other studies have attempted to combine the two types of data, but often keep some parameters fixed in one observational channel while performing a full Markov chain Monte Carlo (NCNC) analysis only for the other \cite{Boltaev:2026prm}. Although these treatments can yield meaningful constraints, they cannot fully characterize the correlations between the intrinsic and additional parameters, nor can they adequately assess the complementary roles of different observational channels in shaping the joint posterior distribution.

Motivated by these considerations, in this work we investigate the Hernquist-type environmental black hole spacetime by combining the EHT measurement of the shadow size of Sgr A* \cite{EventHorizonTelescope:2022wkp}, two publicly available S2 orbital datasets compiled in previous studies \cite{Do:2019txf,Gillessen:2017jxc}, and the Schwarzschild precession factor $f_{\rm sp}$ of the S2 star reported by the GRAVITY Collaboration \cite{GRAVITY:2020gka}. Within a unified Bayesian framework, we perform MCMC sampling using the Python package \texttt{emcee} \cite{Foreman-Mackey:2012any} to investigate the observable effects of the matter surrounding Sgr A* on the black hole spacetime.
Specifically, we first perform separate MCMC analyses using the EHT shadow-size measurement and the S2 orbital data to obtain single-probe constraints on the environmental parameters $(C,\alpha)$. We then construct a joint likelihood function and carry out a joint MCMC analysis to examine the consistency of the constraints provided by the different observational probes and to further constrain the environmental parameters by exploiting their complementary information.

From the perspective of the self-consistency of multiscale gravitational constraints, we investigate whether the nonvacuum metric modifications induced by diffuse environmental matter around Sgr A* can be consistently described by the same spacetime model across the distinct gravitational scales probed by the EHT shadow and the S2 orbit. We then use this cross-scale consistency as a constraint to examine the compatibility of the model parameters across multiple observational scales.
The remainder of this paper is organized as follows. In Sec.\ref{sec:level2}, we introduce the Hernquist-type environmental black hole spacetime and its parametrization. In Sec.\ref{sec:level3}, we describe the observational data, the construction of the likelihood functions, and the Bayesian inference method. In Sec.\ref{sec:level4}, we present the results obtained from the EHT shadow, the S2 orbit, and their joint analysis. 
Finally, Sec. \ref{sec:5} concludes the paper. Throughout this work, all analytical derivations are performed in geometrized units with $G=c=1$.

\section{\label{sec:level2} Hernquist-type environmental black hole spacetime and its parametrization}

Previous models of black hole--environmental matter systems, including those involving dark matter, have mostly employed an additive density prescription within the Newtonian approximation. Such treatments make it difficult to achieve a self-consistent matching between the black hole and its surrounding matter distribution in the strong-field region. To address this limitation, Cardoso et al. \cite{Cardoso:2021wlq} adopted a Hernquist profile for the matter distribution surrounding a black hole at the galactic center and constructed an exact, static, and spherically symmetric solution to the Einstein field equations in the presence of a matter source. In this model, the matter source is described by an anisotropic fluid with the energy-momentum tensor
\begin{equation}
	T^\mu_{\ \nu}=\mathrm{diag}(-\rho,0,P_t,P_t),
\end{equation}
where $\rho$ denotes the energy density, the radial pressure vanishes, and the tangential pressure $P_t$ is determined by the field equations together with the conservation equations. The Hernquist-type density profile motivating the construction is given by \cite{Hernquist:1990be}
\begin{equation}
	\rho=\frac{M a_0}{2\pi r(r+a_0)^3},
\end{equation}
where $M$ is the mass of the matter halo and $a_0$ is its characteristic scale.

Consider a static, spherically symmetric nonvacuum spacetime whose line element can be written as
\begin{equation}
	ds^2=-A(r)dt^2+\frac{dr^2}{B(r)}
	+r^2\left(d\theta^2+\sin^2\theta d\phi^2\right),
	\label{eq:metric}
\end{equation}
where $A(r)=f(r)$, $B(r)=1-2m(r)/r$, and $m(r)$ is the mass function. To simultaneously describe the central black hole and the surrounding distribution of environmental matter, Cardoso et al. \cite{Cardoso:2021wlq} constructed the mass function as
\begin{equation}
	m(r)=M_{\rm BH}
	+\frac{Mr^2}{(a_0+r)^2}
	\left(1-\frac{2M_{\rm BH}}{r}\right)^2,
	\label{eq:mofr}
\end{equation}
where $M_{\rm BH}$ denotes the black hole mass. This construction allows the spacetime to accommodate both the central black hole and the surrounding matter distribution while preserving the self-consistency of the overall geometry.

With this mass function, one obtains an analytical and self-consistent nonvacuum black hole spacetime in general relativity, namely, a black hole spacetime embedded in a Hernquist-type galactic halo. For convenience, we refer to it hereafter as the Hernquist-type environmental black hole spacetime. Its metric function is given by \cite{Cardoso:2021wlq}
\begin{equation}
	f(r)=\left(1-\frac{2M_{\rm BH}}{r}\right)e^{\Upsilon(r)},
	\label{eq:fr}
\end{equation}
where
\begin{equation}
	\Upsilon(r)= -\pi \sqrt{\frac{M}{\xi}}
	+2\sqrt{\frac{M}{\xi}}
	\arctan \left(\frac{r+a_0-M}{\sqrt{M\xi}}\right),
	\label{eq:upsilon}
\end{equation}
and
\begin{equation}
	\xi=2a_0-M+4M_{\rm BH}.
	\label{eq:xi}
\end{equation}
The matter distribution associated with this black hole spacetime is
\begin{equation}
	4\pi\rho
	=\frac{m'(r)}{r^2}
	=\frac{2M(a_0+2M_{\rm BH})}
	{r(r+a_0)^3} \left(1-\frac{2M_{\rm BH}}{r}\right).
\end{equation}
Far from the black hole, this density distribution naturally reduces to the standard Hernquist profile, whereas it vanishes exactly at $r=2M_{\rm BH}$. The event horizon is determined by $g^{rr}=0$ and is located at $r=2M_{\rm BH}$, coinciding with the horizon radius of a Schwarzschild black hole. Therefore, the model self-consistently incorporates environmental effects into the spacetime geometry without shifting the location of the central black hole horizon.

It should be noted that, in addition to the central black hole singularity at $r=0$, the curvature scalar diverges at
$	r=M-a_0\pm\sqrt{M^2-2Ma_0-4MM_{\rm BH}}$
when $M>2(a_0+2M_{\rm BH})$. Therefore, in the following analysis, we impose the condition $M<2(a_0+2M_{\rm BH})$
to exclude additional curvature singularities outside the event horizon and thereby ensure the physical self-consistency of the spacetime under consideration.

To facilitate parameter estimation and comparison with observations, we introduce two dimensionless parameters,
\begin{equation}
	C\equiv \frac{M}{a_0},\qquad
	\alpha\equiv \frac{a_0}{M_{\rm BH}}.
	\label{eq:param}
\end{equation}
Here, $C$ denotes the compactness of the environmental matter and measures how compactly the matter is distributed, while $\alpha$ is the characteristic scale parameter describing the spatial extent of the matter halo. For a typical galactic system, the compactness is generally expected to satisfy $C<10^{-4}$ \cite{Navarro:1995iw}. In the work, however, we consider a galaxy hosting a central black hole and therefore do not impose this restriction, allowing the parameters to vary freely.

In the vicinity of the photon sphere, environmental effects are primarily manifested through modifications to the local geometry and the effective redshift factor. At larger scales, the gravitational potential generated by the environmental matter directly affects stellar orbital dynamics.  
In other words, the EHT observations of the black hole shadow and the S2 star orbital observations do not repeatedly probe the same region, but rather provide complementary information on different radial regions of the same model.
The Hernquist-type environmental black hole spacetime therefore provides a unified metric description of both the near black hole region and the distant galactic environment, offering a rigorous theoretical framework for a multiscale Bayesian analysis combining the black hole shadow data and S2 star orbital data.

\section{\label{sec:level3} Observables and Bayesian inference framework}

In this section, we systematically present the theoretical framework and observational data for the black hole shadow and the orbit of the S2 star, and formulate the likelihood functions, prior distributions, and Bayesian inference framework employed for parameter estimation.

\subsection{Black hole shadow observable}

For the Hernquist-type environmental black hole spacetime described by Eq. \eqref{eq:metric}, the radial equation of motion for photons is given by
\begin{equation}
	\dot{r}^2=B(r)\left(\frac{\mathcal{E}^2}{A(r)}-\frac{\ell^2}{r^2}\right).
\end{equation}
A detailed derivation of the above equation is provided in Appendix \ref{a1}. The boundary of the black hole shadow is determined by unstable circular photon orbits. For the static, spherically symmetric spacetime considered in this work, the photon-sphere radius $r_{\rm ph}$ satisfies
\begin{equation}
	r_{\rm ph} A'(r_{\rm ph}) - 2 A(r_{\rm ph}) = 0.
\end{equation}
Selecting the outer solution that satisfies the instability condition, the corresponding critical impact parameter is
\begin{equation}
	b_c=\frac{r_{\rm ph}}{\sqrt{A(r_{\rm ph})}}.
\end{equation}
For a distant observer, the theoretically predicted angular diameter of the shadow is
\begin{equation}
	\theta_{\rm sh}(M_{\rm BH}, D, C,\alpha)=\frac{2 b_c}{D}.
	\label{eq:shadow}
\end{equation}
where $D$ denotes the distance from the observer to Sgr A*.

We use the effective angular diameter of the shadow defined above to construct the likelihood and constrain the model parameters $\{M_{\rm BH},D,C,\alpha \}$ through a MCMC analysis. Assuming a Gaussian observational uncertainty, the log-likelihood associated with the shadow measurement is
\begin{equation}
	\log \mathcal{L}_{\rm sh}=-\frac{1}{2} \frac{\left(\theta^{the}_{sh}-\theta^{obs}_{sh}\right)^2}{\sigma_\theta^2}.
	\label{eq:lsh}
\end{equation}	
where $\theta_{\rm sh}^{\rm th}$ and $\theta_{\rm sh}^{\rm obs}$ denote the theoretically predicted and observed angular diameters of the shadow, respectively, and $\sigma_\theta$ is the corresponding observational uncertainty.

\subsection{S2 star observables}

The primary observables of the S2 star are its relative angular position $(\alpha,\delta)$ on the plane of the sky and its line-of-sight velocity $RV$. The theoretical orbit is solved in the equatorial plane, whereas the observational data are defined in an observer frame formed by the tangent plane to the celestial sphere and the line-of-sight direction.

Within the framework of general relativity, the orbital equations governing the motion of the S2 star in the equatorial plane of the central Hernquist-type environmental black hole spacetime can be written as
\begin{equation}
	\frac{d^2u}{d\phi^2} = F(u), \qquad \frac{d\phi}{d t}=\frac{A_u \ell u^2  }{\mathcal{E}}, \qquad u = \frac{1}{r}.
\end{equation}
A detailed derivation is provided in Appendix \ref{a1}. In principle, the starting point of the orbital integration may be chosen arbitrarily, provided that the resulting trajectory is subsequently aligned with the observational epochs. In this work, the apocenter is adopted as the reference point, and the orbit is evolved both forward and backward in time from this point. The corresponding initial conditions are
\begin{equation}
	t_0=t_{\text{ref}},\; r(t_0)=a (1+e),\; \phi(t_0)=\pi ,\; \dot{r}(t_0)=0.
\end{equation}
The numerical integration yields
\begin{equation}
r(t),\quad \phi(t),\quad \dot{r}(t),\quad \dot{\phi}(t).
\end{equation}

In the orbital plane, the position and velocity vectors are respectively
given by
\begin{equation}
	\mathbf{r}_{\rm orb}(t)=
	\begin{bmatrix}
		x\\
		y\\
		z
	\end{bmatrix}=
	\begin{bmatrix}
		r(t)\cos\phi(t)\\
		r(t)\sin\phi(t)\\
		0
	\end{bmatrix},
\end{equation}
and
\begin{equation}
	\mathbf{v}_{\rm orb}(t)=
	\begin{bmatrix}
		v_x\\
		v_y\\
		v_z
	\end{bmatrix}=
	\begin{bmatrix}
		\dot{r}(t)\cos\phi(t)
		-r(t)\dot{\phi}(t)\sin\phi(t)
		\\
		\dot{r}(t)\sin\phi(t)
		+r(t)\dot{\phi}(t)\cos\phi(t)
		\\
		0
	\end{bmatrix}.
\end{equation}
Here, $\dot{r}(t)$ and $\dot{\phi}(t)$ denote derivatives with respect to the coordinate time $t$.

Before comparison with the observational data, the coordinates in the orbital plane must be transformed into the observer frame.The corresponding Euler rotation transformation is given by
\begin{equation}
\mathbf{R}_{E}= R_3(\Omega)R_1(i)R_3(\omega),
\end{equation}
where $R_1(\cdot)$ and $R_3(\cdot)$ denote rotation matrices about the first and third coordinate axes, respectively. The angles $\Omega$, $i$, and $\omega$ are the position angle of the ascending node, the orbital inclination, and the argument of pericenter, respectively.

To ensure that the transformed coordinate components are consistent with the axis convention of the observational frame, we introduce the coordinate alignment matrix
\begin{equation}
	\mathbf{P}_{\rm obs}=
	\begin{bmatrix}
		0 & 1 & 0\\
		1 & 0 & 0\\
		0 & 0 & 1
	\end{bmatrix}.
\end{equation}
This matrix only rearranges the coordinate components to match the adopted observational convention and does not represent an additional physical rotation of the orbit. The complete transformation from the orbital plane to the observer frame is therefore
\begin{equation}
\mathbf{R}_{\rm obs}=\mathbf{P}_{\rm obs}\mathbf{R}_{E}=\mathbf{P}_{\rm obs} R_3(\Omega)R_1(i)R_3(\omega).
\end{equation}
The position and velocity vectors in the observer frame are then
\begin{align}
\mathbf{r}_{\rm obs}(t)=\mathbf{R}_{\rm obs}\mathbf{r}_{\rm orb}(t),\quad
\mathbf{v}_{\rm obs}(t)=\mathbf{R}_{\rm obs}\mathbf{v}_{\rm orb}(t),
\end{align}
where
\begin{align}
\mathbf{r}_{\rm obs}(t)=&\bigl[X(t),Y(t),Z(t)\bigr]^{\mathsf{T}}, \notag \\
\mathbf{v}_{\rm obs}(t)=&\bigl[V_X(t),V_Y(t),V_Z(t)\bigr]^{\mathsf{T}}.
\end{align}
Here, $X(t)$ and $Y(t)$ are the two coordinate components on the plane of the
sky and correspond, respectively, to the relative right-ascension and relative declination directions adopted in this work. The coordinate $Z(t)$ is the component along the line of sight.

Because astrometric measurements provide angular positions on the plane of the sky, only the $X$ and $Y$ components are required when comparing the model with the observations. The theoretically predicted angular positions
are written as \cite{Do:2019txf}
\begin{equation}
	\begin{bmatrix}
		\alpha_{\rm th}(t_{\rm obs})\\
		\delta_{\rm th}(t_{\rm obs})
	\end{bmatrix}
	=
	\frac{1}{D}
	\begin{bmatrix}
		X(t_{\rm em})\\
		Y(t_{\rm em})
	\end{bmatrix}
	+
	\begin{bmatrix}
		x_0\\
		y_0
	\end{bmatrix}
	+
	\begin{bmatrix}
		v_{x0}\\
		v_{y0}
	\end{bmatrix}
	\left(t_{\rm obs}-t_{\rm ref}\right).
\end{equation}
Here, $D$ is the distance from the observer to the Galactic center; $(x_0,y_0)$ are the astrometric zero-point offsets; and $(v_{x0},v_{y0})$ describe the linear drift of the astrometric reference frame on the plane of the sky. The reference epoch is denoted by
$t_{\rm ref}$, for which we adopt $t_{\rm ref}=2009.2$. The quantities $t_{\rm em}$ and $t_{\rm obs}$ denote the photon emission and observation times, respectively. These two epochs differ because of the finite propagation time between the emission of a photon by the S2 star and its detection by the observer. Consequently, the observational epoch does not coincide with the corresponding emission epoch, and the R{\o}mer time delay must be included consistently \cite{AIHPA_1986__44_3_263_0,GRAVITY:2018ofz}:
\begin{equation}
t_{\rm em}-t_{\rm obs}= \frac{Z(t_{\rm em})}{c}.
\end{equation}

In addition to positional observations of S2's orbit, the radial velocity $RV$ is also obtained through spectroscopic observations. The frequency shift of the spectrum is defined as
\begin{equation}
\zeta = \frac{\Delta\nu}{\nu} =\frac{\nu_{\rm em}-\nu_{\rm obs}}{\nu_{\rm obs}}=\frac{RV}{c},
\end{equation}
where $\nu_{\mathrm{em}}$ is the emission frequency, $\nu_{\mathrm{obs}}$ is the observed frequency, and $RV$ denotes the radial velocity \cite{QiQi:2026pnb}. This frequency shift comprises two independent physical contributions. The first is the relativistic Doppler shift, which primarily arises from the high‑speed motion of the S2 star, and takes the form
\begin{equation}
\zeta_D=\frac{\sqrt{1-v^2(t_{\rm em})/c^2}}{1-\hat{\mathbf{n}}\cdot\mathbf{v}(t_{\rm em})/c
}.
\end{equation}
where $\hat{n}\cdot\vec{v}(t_{\mathrm{em}}) = V_Z$ represents the projection of the space velocity along the line of sight \cite{DellaMonica:2021xcf}.  
The second is the gravitational redshift induced by the strong gravitational field, expressed as
\begin{equation}
\zeta_G =\frac{1}{\sqrt{-g_{00}\bigl(t_{\rm em},\mathbf{r}_{\rm em}\bigr)}}=\frac{1}{\sqrt{A\bigl(r_{\rm em}\bigr)}}.
\end{equation}
Combining the dual contributions of the Doppler shift and the gravitational redshift, the theoretical total frequency shift factor is expressed as
\begin{equation}
	1 + \zeta = \zeta_D \cdot \zeta_G.
\end{equation}
Converting the total frequency shift into a radial velocity and introducing the systemic radial velocity calibration term $v_{\mathrm{LSR}}$, the theoretical radial velocity can be written as \cite{Do:2019txf,Yao:2026pjz}
\begin{equation}
RV_{the}= c \cdot\zeta +v_{LSR}.
\label{RV}
\end{equation}

Combining the above analysis, the theoretical predictions of the model considered in this paper can be fitted to actual astronomical observations. The specific form is given by 
\begin{widetext}
\begin{align}
\underbrace{(a,e,M_{\rm BH},D,i,\Omega,\omega,x_0,y_0,v_{x0},v_{y0},\delta t_p,v_{LSR},C,\alpha)}_{\text{model parameters}} \Rightarrow \underbrace{(\alpha_{the},\delta_{the},RV_{the})\Leftrightarrow(\alpha_{obs},\delta_{obs},RV_{obs})}_{\text{Model predictions vs. observations}}
\end{align}
\end{widetext}
On the left‑hand side, $\{M_{\mathrm{BH}}, D\}$ denote the mass of Sgr A* and its distance from the observer, respectively. The set $\{a, e, i, \Omega, \omega\}$ comprises the orbital elements of the S2 star, while $\{x_0, y_0, v_{x0}, v_{y0}, v_{\mathrm{LSR}}\}$ includes the reference‑frame offsets, drift terms, and the systemic line‑of‑sight velocity offset. $\delta t_p$ denotes the offset of the S2 star’s pericenter passage time relative to the reference epoch of 2002, i.e., $t_p - 2002$; $\{C, \alpha\}$ represent the parameters of the environmental matter in a Hernquist‑type black hole spacetime.
The right-hand side of the expression represents the theoretically predicted right ascension, declination, and radial velocity, in comparison with the corresponding observed values.
Based on this parameter mapping, a Bayesian inference on the model parameters can be performed using the MCMC method combined with the observational data.

The model likelihood function can be composed of the astrometric term, the radial velocity term, and the orbital precession term 
\begin{equation}
	\log \mathcal{L}_{\rm S2} = \log \mathcal{L}_{\rm P} + \log \mathcal{L}_{\rm RV} + \log \mathcal{L}_{\rm pre},
\end{equation}
where the astrometric position likelihood term $\log \mathcal{L}_{\rm P}$ is constructed from the right‑ascension and declination offsets of S2 star relative to Sgr A*, and is given by
\begin{equation}
	\log \mathcal{L}_{\rm P} = -\frac{1}{2} \sum_{P}^i \left[ \left(\frac{\alpha_{\rm the}^i - \alpha_{\rm obs}^i}{\sqrt{2}\,\sigma_\alpha^i}\right)^2 + \left(\frac{\delta_{\rm the}^i - \delta_{\rm obs}^i}{\sqrt{2}\,\sigma_\delta^i}\right)^2 \right].
\end{equation}
The radial velocity likelihood term $\log\mathcal{L}_{\rm RV}$ is constructed from the radial velocity observations of S2 star, and is given by
\begin{equation}
	\log \mathcal{L}_{\rm RV} = -\frac{1}{2} \sum_{RV}^j \left(\frac{RV_{\rm the}^j - RV_{\rm obs}^j}{\sqrt{2}\,\sigma_{RV}^j}\right)^2.
\end{equation}
The orbital precession likelihood term $\log \mathcal{L}_{\rm pre}$ is constructed from the observational constraints on S2 star's Schwarzschild precession parameter $f_{\rm sp}$, and is given by
\begin{equation}
	\log \mathcal{L}_{\rm pre} = -\frac{1}{2} \left(\frac{f_{\rm sp,\,the} - f_{\rm sp,\,obs}}{\sqrt{2}\,\sigma_{\rm sp}}\right)^2.
\end{equation}
In the above equations, the subscript ``the'' denotes the theoretical prediction of the model, the subscript ``obs'' denotes the observed value, and $\sigma$ is the uncertainty of the corresponding measurement. The weight factor $\sqrt{2}$ in each term arises because the precession information and the orbital data come from the same set of observations, causing that data to be effectively counted twice. This factor is introduced precisely to compensate for the overweighting that would otherwise result from this double counting \cite{Yao:2026pjz,DellaMonica:2021xcf,Bambhaniya:2026slo}.

\subsection{Joint Likelihood and Parameter Priors}

EHT shadow angular diameter observations primarily constrain the apparent shadow size of Sgr A* near the event horizon scale, and are sensitive to the spacetime structure in the near‑horizon region. The positional and radial velocity data of the S2 star probe the gravitational field on the stellar orbital scale. Since the two types of data correspond to different observables and their measurement error sources are independent of each other, a joint analysis can provide complementary constraints on the black hole spacetime at different radial scales and improve the ability to constrain the model parameters. The joint likelihood can be written as the product of the individual likelihoods, i.e.,
\begin{equation}
	\mathcal{L}_{\rm joint} = \mathcal{L}_{\rm sh} \cdot \mathcal{L}_{\rm S2},
\end{equation}
which is equivalent to
\begin{equation}
	\log \mathcal{L}_{\rm joint} = \log \mathcal{L}_{\rm sh} + \log \mathcal{L}_{\rm S2}.
\end{equation}

For the Hernquist‑type environmental black hole spacetime described by the line element \eqref{eq:metric}, the non‑vacuum modification in the metric is characterized by the compactness $C$ and the characteristic scale $\alpha$ in Eq. \eqref{eq:param}. We first perform single‑probe MCMC parameter inference based on the black hole shadow data and the S2 star data separately. For the black hole shadow observations, the set of sampled parameters is
\begin{equation}
\theta_{\rm sh}=\{M_{\rm BH},D,C,\alpha\}.
\end{equation}

For the orbital observations of the S2 star, in addition to the aforementioned shared parameters, the stellar orbital parameters as well as the zero‑point offset parameters in the astrometric and radial velocity observations also need to be sampled simultaneously. Therefore, the parameter set used in the analysis of the S2 star observations is
\begin{equation}
\theta_{\rm S2}=\{a,e,M_{\rm BH},D,i,\Omega,\omega,x_0,y_0,v_{x0},v_{y0},\delta t_p,v_{\rm LSR},C,\alpha\}.
\end{equation}

In the joint analysis, the posterior distribution is constructed in a unified parameter space. Since the parameters on which the black hole shadow likelihood depends, $\{M_{\rm BH}, D, C, \alpha\}$, are already included in the parameter set $\theta_{\mathrm{S2}}$ of the S2 star data, the sampling parameter set for the joint analysis can be written as
\begin{equation}
\theta_{\rm joint}=\theta_{\rm S2}.
\end{equation}

To ensure comparability between different observational constraints, this paper adopts identical prior distributions for the shared parameters in the Bayesian inference using the black hole shadow data, the S2 star data, and their joint data. Tab. \ref{tab1} lists the prior distributions of each sampled parameter in the three types of analysis. With the exception of the environmental parameters $C$ and $\alpha$, which adopt log‑uniform priors, all other parameters adopt uniform priors. The log‑uniform priors correspond to parameter values spanning several orders of magnitude, covering a broad parameter space from the extremely weak environment to the strongly compact environment, and from the near‑black‑hole scale to the galactic scale, so as to assess the influence of the model on the observables under different environmental conditions.

\begin{table}[htbp]
\centering
\renewcommand{\arraystretch}{1.2}
\caption{Prior distributions of the model parameters, where $\mathcal{U}(a,b)$ denotes a uniform distribution with lower bound $a$ and upper bound $b$.}
\label{tab1}
\begin{tabular}{p{2.5cm}p{3.5cm}p{2cm}}
\hline\hline
Parameter & Prior distribution & Unit \\
\hline
$a$ & $\mathcal{U}(110,\;140)$ & $mas$ \\
$e$ & $\mathcal{U}(0.8,\;0.95)$ & --- \\
$M_{\rm BH}$ & $\mathcal{U}(2,\;6)$ & $\times 10^6 M_\odot$ \\
$D$ &  $\mathcal{U}(6,\;10)$ & $kpc$ \\
$i$ & $\mathcal{U}(86,\; 172)$ & $^\circ$ \\
$\Omega$ & $\mathcal{U}(200,\;258)$ & $^\circ$ \\
$\omega$ & $\mathcal{U}(29,\;86)$ & $^\circ$ \\
$x_0$ & $\mathcal{U}(-5,\;5)$ & $mas$ \\
$y_0$ & $\mathcal{U}(-5,\;5)$ & $mas$ \\
$v_{x0}$ & $\mathcal{U}(-5,\;5)$ & $mas/yr$ \\
$v_{y0}$ & $\mathcal{U}(-5,\;5)$ & $mas/yr$ \\
$t_p-2002$ & $\mathcal{U}(-1,\;1)$ & $yr$ \\
$v_{LSR}$ & $\mathcal{U}(-100,\;100)$ & $km/s$ \\
$\log_{10}(C)$ & $\mathcal{U}(-8,\;0)$ & --- \\
$\log_{10}(\alpha)$ & $\mathcal{U}(1,\;11)$ & --- \\
\hline\hline
	\end{tabular}
\end{table}

\subsection{Observational Data}

In performing MCMC sampling to constrain the model parameters, the observational data used in this work come from black hole shadow observations and S2 star observations. Specifically, for the shadow observational data, we adopt the effective geometric shadow angular diameter $\theta_{\rm sh}^{\rm obs} = (48.7 \pm 7)~\mu$as given by the EHT Collaboration based on images of Sgr A* and numerical simulation calibrations \cite{EventHorizonTelescope:2022wkp}.
For the S2 star orbital data, we adopt the data compiled by Do et al.\ \cite{Do:2019txf} (Dataset 1) and the data compiled by Gillessen et al.\ \cite{Gillessen:2017jxc} (Dataset 2). In the analysis of both datasets, the precession data from the GRAVITY Collaboration \cite{GRAVITY:2020gka} are also incorporated. The precession parameter is given by
\begin{equation}
	f_{\rm sp} = \frac{\Delta\phi}{\Delta\phi_{\rm GR}} = 1.1 \pm 0.19.
\end{equation}
Dataset 1 contains 46 positional measurements $(\alpha_{\rm obs},\delta_{\rm obs})$  from 1995 to 2018 and 116 radial velocity measurements from 2000 to 2019. Dataset 2 contains 145 positional measurements from 1992 to 2016 and 44 radial velocity measurements $RV_{\rm obs}$.

\section{\label{sec:level4}Results}

\begin{figure}[htbp]
	\centering
	\includegraphics[width=0.45\textwidth]{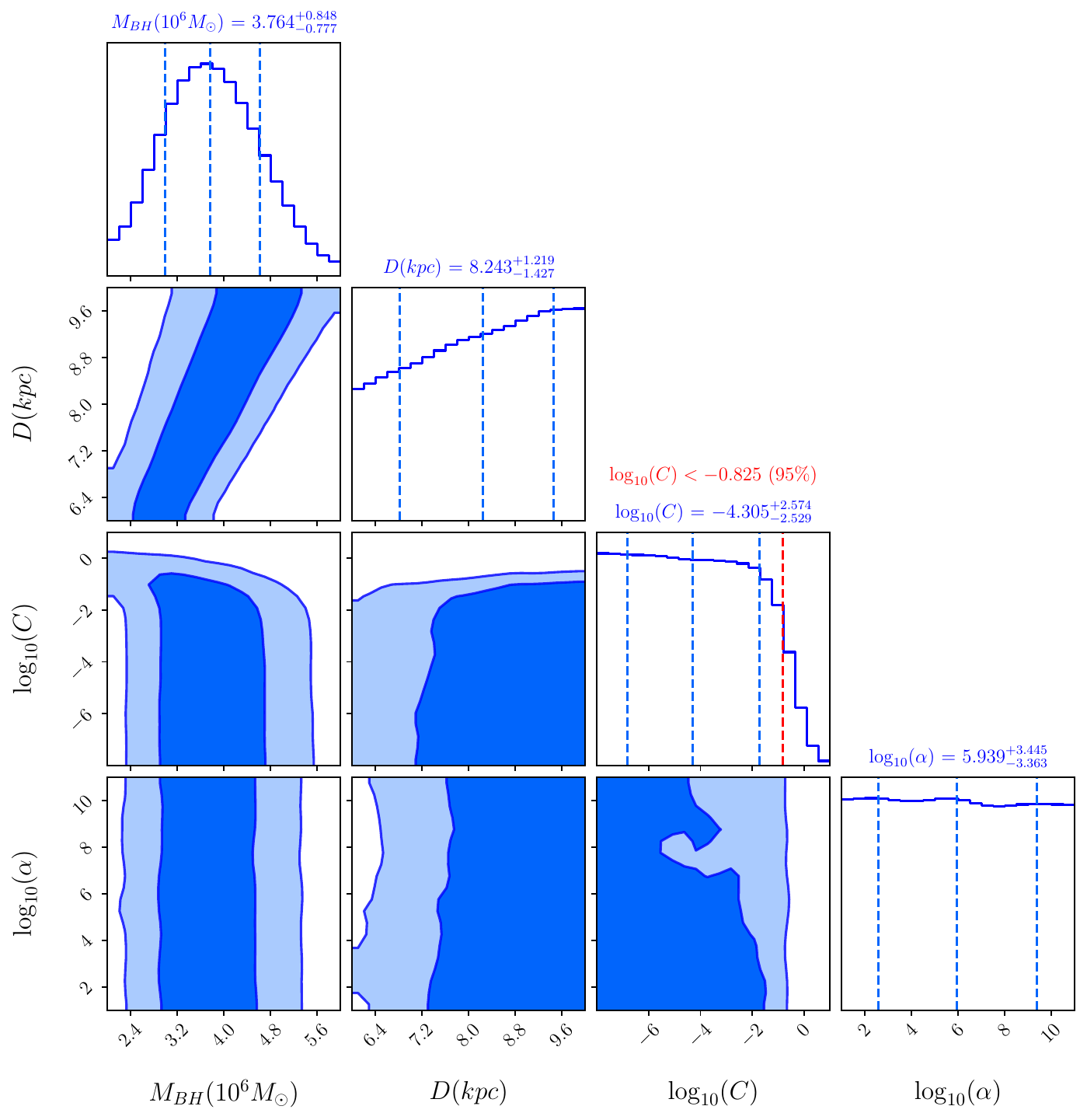}
	\caption{Posterior distributions of the parameters $\{M_{\mathrm{BH}}, D, C, \alpha\}$ obtained from the EHT observational data. The two‑dimensional contour lines correspond to the 68\% and 95\% confidence intervals; in the one‑dimensional histograms, the blue dashed lines indicate the median and the $1\sigma$ confidence intervals, and the red dashed line marks the 95\% credible upper limit of $\log_{10}(C)$.}
	\label{fig:corner1}
\end{figure}

\begin{figure*}[htbp]
	\centering
	\includegraphics[width=1\textwidth]{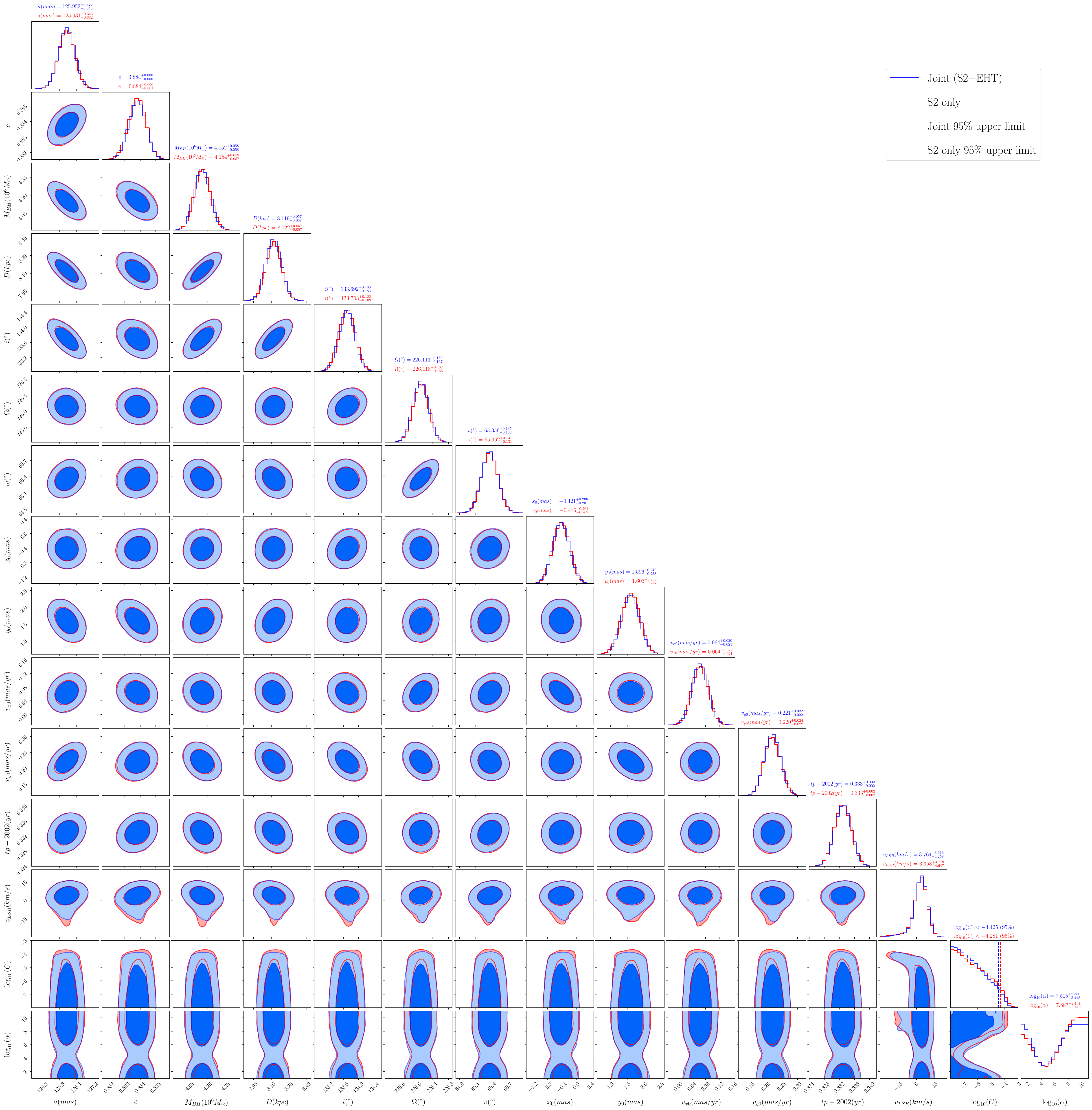}
	\caption{15‑dimensional posterior distributions of the parameters obtained from the S2 star data alone (red) and from the joint S2+EHT data (blue). The two‑dimensional contours correspond to the 68\% and 95\% confidence intervals; the one‑dimensional histograms show the posterior median and the $1\sigma$ confidence intervals for each parameter, and the dashed line indicates the 95\% credible upper limit of $\log_{10}(C)$. The S2 star observational data are from Dataset 1.}
	\label{fig:corner2}
\end{figure*}

\begin{figure*}[htbp]
	\centering
	\includegraphics[width=1\textwidth]{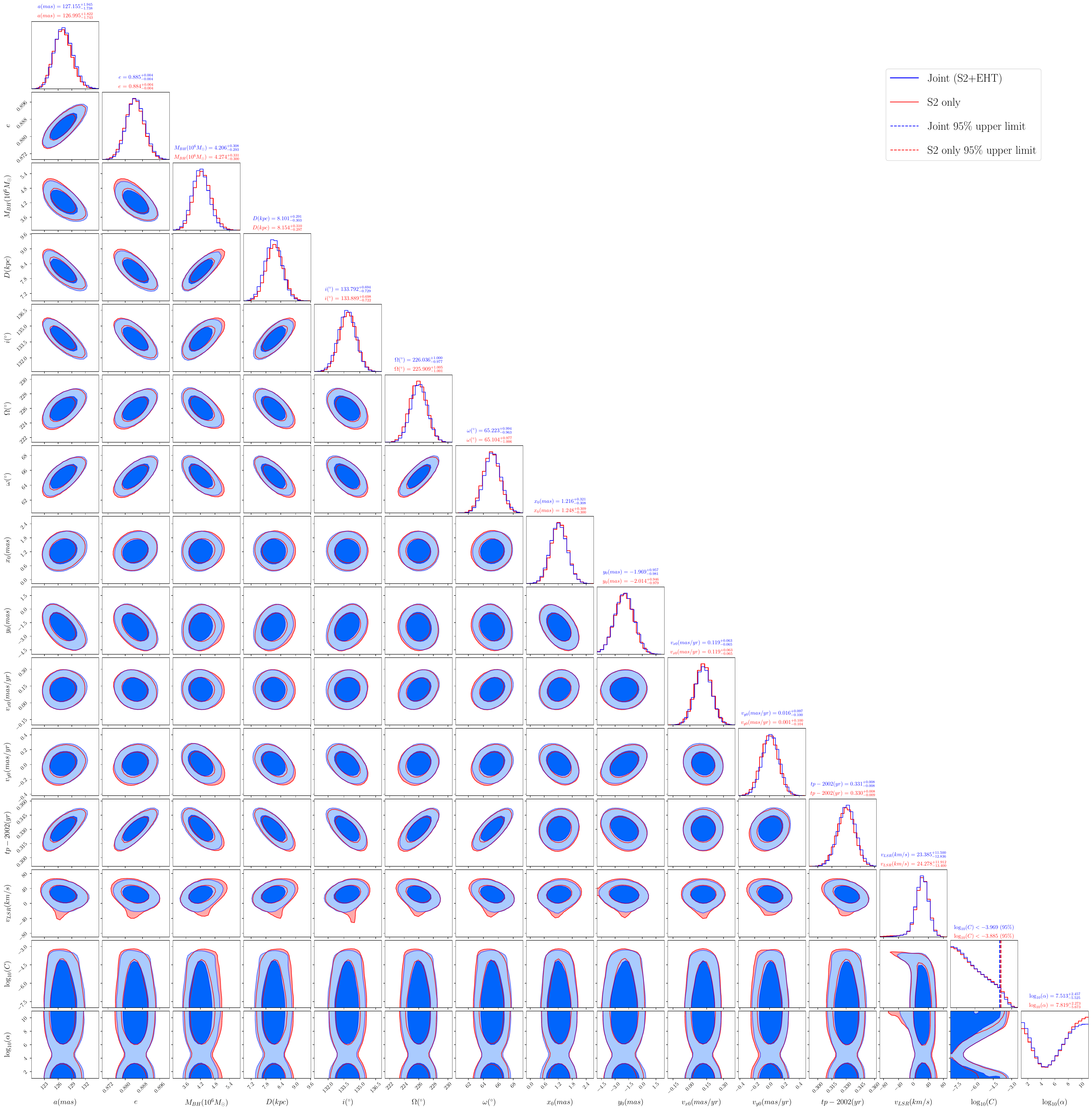}
	\caption{Same as Fig. \ref{fig:corner2}, but with the S2 star observational data from Dataset 2.}
	\label{fig:corner3}
\end{figure*}

We perform the MCMC analysis using the Python package \texttt{emcee}  \cite{Foreman-Mackey:2012any} and produce corner plots using the \texttt{corner} package \cite{corner}. 
In the sampling based solely on the shadow data, 16 walkers are run in parallel; in the sampling based on the S2 star data and in the joint sampling, 40 walkers are run in parallel.
Tab. \ref{tab2} summarizes the posterior medians and their $1\sigma$ confidence intervals for each parameter obtained from the MCMC analysis, where the compactness  $C$ is given as a 95\% credible upper limit. The corresponding corner plots of the posterior distributions are shown in Figs. \ref{fig:corner1}--\ref{fig:corner3}.

\begin{table*}[htbp]
	\centering
  \renewcommand{\arraystretch}{1.3}
	\caption{Posterior distributions of model parameters. For each parameter, the median and $1\sigma$ credible interval are listed; for the compactness $C$, the 95\% credible upper limit is listed. Dataset 1 is from Do et al. \cite{Do:2019txf}, and Dataset 2 is from Gillessen et al. \cite{Gillessen:2017jxc}.}
	\label{tab2}
	\begin{tabular}{p{2.6cm}p{2.6cm}p{2.6cm}p{2.6cm}p{0.2cm}p{2.6cm}p{2.6cm}}
		\hline\hline
		& & \multicolumn{2}{c}{Dataset 1} & & \multicolumn{2}{c}{Dataset 2} \\
		\cline{3-4} \cline{6-7}
		Parameter & EHT  & S2 only & joint(S2+EHT) & & S2 only & joint(S2+EHT) \\
		\hline
		$a(mas)$              & ---          &  $125.931^{+0.343}_{-0.332}$ &$125.952^{+0.329}_{-0.340}$ & &$126.995^{+1.822}_{-1.743}$&$127.155^{+1.945}_{-1.738}$\\
		$e$                  & ---            &  $0.884^{+0.000}_{-0.001}$    &$0.884^{+0.000}_{-0.000}$ & &$0.884^{+0.004}_{-0.004}$&$0.885^{+0.004}_{-0.004}$\\
		$M_{\rm BH}(10^6 M_\odot)$ & $3.764^{+0.848}_{-0.777}$  & $4.154^{+0.059}_{-0.057}$ &$4.152^{+0.058}_{-0.058}$ & &$4.274^{+0.331}_{-0.300}$ &$4.206^{+0.308}_{-0.293}$\\
		$D(kpc)$ & $8.243^{+1.219}_{-1.427}$  & $8.122^{+0.057}_{-0.057}$ & $8.119^{+0.057}_{-0.057}$ & & $8.154^{+0.310}_{-0.297}$&$8.101^{+0.291}_{-0.303}$\\
		$i(^\circ)$ & --- & $133.703^{+0.188}_{-0.189}$&$133.692^{+0.183}_{-0.185}$&&$133.889^{+0.698}_{-0.722}$ &$133.792^{+0.694}_{-0.729}$\\
		$\Omega(^\circ)$ & --- &$226.118^{+0.167}_{-0.165}$&$226.113^{+0.163}_{-0.167}$&&$225.909^{+1.005}_{-1.001}$&$226.036^{+1.000}_{-0.977}$\\
		$\omega(^\circ)$ & --- & $65.362^{+0.131}_{-0.131}$&$65.359^{+0.135}_{-0.133}$&&$65.104^{+0.977}_{-1.006}$ &$65.223^{+0.994}_{-0.963}$\\
		$x_0(mas)$ & --- &$-0.416^{+0.201}_{-0.202}$&$-0.421^{+0.206}_{-0.201}$ &&$1.248^{+0.309}_{-0.300}$&$1.216^{+0.321}_{-0.308}$\\
		$y_0(mas)$ &--- & $1.603^{+0.238}_{-0.247}$&$1.596^{+0.242}_{-0.238}$&&$-2.014^{+0.946}_{-0.979}$&$-1.969^{+0.957}_{-0.981}$\\
		$v_{x0}(mas/yr)$ & --- & $0.064^{+0.022}_{-0.021}$&$0.064^{+0.020}_{-0.021}$&&$0.119^{+0.063}_{-0.065}$&$0.119^{+0.063}_{-0.065}$\\
		$v_{y0}(mas/yr)$ & --- &$0.220^{+0.024}_{-0.022}$&$0.221^{+0.022}_{-0.022}$&&$0.001^{+0.100}_{-0.104}$ &$0.016^{+0.097}_{-0.100}$\\
		$t_p-2002(yr)$ & --- & $0.333^{+0.002}_{-0.002}$&$0.333^{+0.002}_{-0.002}$&&$0.330^{+0.008}_{-0.009}$&$0.331^{+0.008}_{-0.008}$\\
		$v_{LSR}(km/s)$ &--- & $3.353^{+3.718}_{-4.647}$&$3.764^{+3.613}_{-4.258}$&&$24.278^{+11.912}_{-13.400}$ &$23.385^{+11.500}_{-12.836}$\\
		$\log_{10} (C)$ & $\lesssim-0.825(95\%)$ & $\lesssim-4.281(95\%)$& $\lesssim-4.425(95\%)$ & & $\lesssim-3.885(95\%)$& $\lesssim-3.969(95\%)$\\
		$\log_{10}(\alpha)$ & $5.939^{+3.445}_{-3.363}$ & $7.887^{+2.137}_{-5.440}$&$7.515^{+2.380}_{-5.415}$&&$7.819^{+2.273}_{-5.673}$&$7.513^{+2.457}_{-5.525}$ \\
		\hline\hline
	\end{tabular}
\end{table*}

Fig. \ref{fig:corner1} shows the posterior distributions of the parameters obtained using the EHT measured shadow angular diameter as the observational data. The contour lines in the figure correspond to the 68\% and 95\% confidence intervals, and the red dashed line indicates the 95\% credible upper limit. Under broad uniform priors, the 95\% credible upper limit of the compactness  $C$ is $\log_{10} (C) \lesssim -0.825$ (corresponding to $C \lesssim 1.498 \times 10^{-1}$), which is consistent with the analytical upper limit result obtained in Feng et al.\ \cite{Feng:2025iao} with $M$ and $D$ fixed. In addition, the characteristic scale $\alpha$ is not substantially constrained, and the posteriors of $M_{\rm BH}$ and $D$ exhibit a positive correlation.
Fig. \ref{fig:corner2} shows an overlay comparison of the posterior distributions of the parameters obtained from the S2 star data alone (red) and from the joint S2+EHT data (blue). The S2 star observational data are from Dataset 1. The S2 star data alone yield a 95\% credible upper limit of $\log_{10}(C) \lesssim -4.281$ (corresponding to $C \lesssim 5.239 \times 10^{-5}$). The joint data further tighten this constraint to $\log_{10}(C) \lesssim -4.425$ (corresponding to $C \lesssim 3.760 \times 10^{-5}$), and all model parameters are further narrowed. For $\alpha$, a bimodal structure is observed under both analyses, with the two peaks located at the two boundaries of the parameter space. The physical origin of this feature will be discussed later.
Fig. \ref{fig:corner3} has the same layout as Fig. \ref{fig:corner2}, but the S2 star observational data use Dataset 2.  
Analysis using only the S2 star data yields a 95\% credible upper limit of $\log_{10}(C) \lesssim -3.885$ (corresponding to $C \lesssim 1.303 \times 10^{-4}$), while the joint observational data analysis tightens it to $\log_{10}(C) \lesssim -3.969$ (corresponding to $C \lesssim 1.073 \times 10^{-4}$). The bimodal positions of $\alpha$ are consistent with those in Fig. \ref{fig:corner2}. Furthermore, the posterior distributions of the radial velocity offset $v_{\rm LSR}$ for both datasets exhibit a non-Gaussian negative tail. This originates from the contribution of a constant gravitational redshift term, $C \cdot c$, from the time component $g_{00}$ of the metric in Eq. (\ref{RV}), which introduces a linear compensation effect in the parameter fitting.  Overall, the parameter constraints derived from Dataset 2 are weaker than those from Dataset 1.

The bimodal structure of $\alpha$ originates from the difference in the constraining power of the environmental potential $\Phi_{\rm env} \approx C \alpha M_{\rm BH}(2 M_{\rm BH}-r)/(r\alpha M_{\rm BH}+r^2)$.
When $\alpha M_{\rm BH} \ll r_{\mathrm{S2}}$, the S2 star's orbit lies outside the environmental matter halo; in this regime, the halo can be treated as a central point mass, and its effect is degenerate with the central black hole mass. When $\alpha M_{\rm BH} \gg r_{\rm S2}$, the S2 star's orbit lies well inside the halo; the environmental potential is approximately constant and does not perturb the orbital motion. In these two limits, the environmental matter halo fails to produce significant orbital perturbations, so the allowed parameter space for $C$ is broad, forming a bimodal structure at the two extremes of $\alpha$.The valley floor between the two peaks appears near $\alpha \sim 10^4$. At this parameter, the characteristic scale of the environmental matter halo lies exactly between the pericenter and apocenter of the S2 star's orbit, causing the difference in the gravitational field produced by the environmental matter between these two points to reach its maximum. This difference  affects the precession of the S2 star's orbit, thereby strongly compressing the allowed parameter space for $C$. We also independently constrain $C$ and $\alpha$ in Appendix \ref{B} using the precession data from the GRAVITY collaboration \cite{GRAVITY:2020gka}, and the results further corroborate the physical nature of the bimodality.
It is worth noting that when $\alpha M_{\rm BH} \ll r_{\mathrm{S2}}$, the gravitational contribution of the environmental matter can be equivalent to a point mass $M_{\rm env} \sim C\alpha M_{\rm BH}$. In our parameter sampling, taking $\alpha M_{\rm BH} \sim 10\,M_{\rm BH}$, the compactness $C$ can be as large as $C \sim 10^{-4}$, and the corresponding equivalent mass is $M_{\rm env} \sim 10^{-3}\,M_{\rm BH}$. This result is consistent with the conclusion of the GRAVITY collaboration that the extended mass within the S2 star's orbit does not exceed one thousandth of the central black hole mass \cite{GRAVITY:2020gka,GRAVITY:2021xju}.

In summary, compared to using only the shadow diameter data, the S2 star observational data not only significantly tighten the constraint on the compactness $C$, but also provide meaningful constraints on the characteristic scale $\alpha$, with its posterior marginal distribution exhibiting a bimodal structure. Analysis of the joint data further tightens all model parameters; the specific results are given in Tab. \ref{tab2}. The stringent upper limit on $C$ rules out the possibility of a compact matter environment surrounding the central black hole in the galactic center, and this limit is consistent with the typical compactness range $C<10^{-4}$ required by galactic halo structures \cite{Navarro:1995iw}. These results demonstrate the unique value of multi-probe joint analysis in testing gravitational theories and constraining the black hole environment.
In the future, with improved observational precision and extended observation time span, joint analysis of multiple stars is expected to further identify the actual branch of the $\alpha$ bimodality and tighten the upper limit on $C$.

\section{\label{sec:5}Discussion and conclusions}

In this paper, we perform Bayesian inference on the model parameters using the MCMC method and systematically investigate the Hernquist-type environmental black hole spacetime. This spacetime is an exact non-vacuum solution to the Einstein field equations with an anisotropic matter source, where the environmental matter distribution asymptotically approaches a Hernquist profile on large scales. It thus provides a self-consistent fully relativistic framework for studying black hole environmental effects in strong gravitational fields. Based on this model, we numerically solve null and timelike geodesics, and combine the black hole shadow  data of Sgr A* with the astrometric and radial velocity data of the S2 star. Through separate and joint MCMC analyses, we constrain the compactness $C$ and the characteristic scale $\alpha$ that describe the environmental matter.

Our results show that when performing MCMC analysis using only the black hole shadow size data, the 95\% credible upper limit on the environmental matter compactness $C$ is $C\lesssim 1.498\times10^{-1}$, while the characteristic scale $\alpha$ is not meaningfully constrained, indicating that the current shadow size data have low sensitivity to $\alpha$. The constraining power improves substantially when using only the S2 star data. Specifically, using the dataset of Do et al.\cite{Do:2019txf} yields $C\lesssim 5.239\times10^{-5}$ (95\%), and using the dataset of Gillessen et al.\cite{Gillessen:2017jxc} yields $C\lesssim 1.303\times10^{-4}$ (95\%), with the posterior distributions of $\alpha$ both exhibiting a clear bimodal structure.
In comparison, the constraints from the Do et al. dataset are tighter than those from the Gillessen et al. dataset, because the former includes high-precision positional and radial velocity measurements from 2016--2019, thus providing stronger constraining power.
The appearance of the $\alpha$ bimodality indicates that, at the current observational precision, there exists a parameter degeneracy in the characteristic scale, which cannot be fully broken by the limited radial range covered by the S2 star alone. When the shadow and S2 star data are further combined, the posterior distributions of all parameters contract. In particular, the 95\% credible upper limits on $C$ are tightened to $C\lesssim 3.760\times10^{-5}$ (Do et al. dataset) and $C\lesssim 1.073\times10^{-4}$ (Gillessen et al. dataset), respectively. However, the bimodal structure does not disappear; only its relative weights undergo minor changes. This indicates that the degeneracy is not a data problem, but arises because two different parameter combinations become observationally indistinguishable due to the equivalent effect of the environmental potential. When $\alpha M_{\rm BH}$ is much smaller than or much larger than the orbital radius of the S2 star, the environmental potential produces nearly equivalent effects on the stellar motion, thereby causing the degeneracy. When $\alpha M_{\rm BH}$ is comparable to the S2 orbital radius, the gradient of the environmental potential varies significantly, and its influence on the stellar motion is most prominent. In this regime, only a lower compactness $C$ can be compatible with the current observational data.

In summary, we adopt the Hernquist-type environmental black hole spacetime to describe Sgr A* and, within a Bayesian framework, employ the MCMC method to constrain the compactness $C$ and the characteristic scale $\alpha$ of the environmental matter. The results restrict the compactness $C$ of the environmental matter around Sgr A* to be below the order of $10^{-5}$--$10^{-4}$, consistent with typical galaxy halo density limits \cite{Navarro:1995iw}. This rules out the existence of compact environmental matter, while a low-density environment with weak gravitational contribution remains compatible with observations. In contrast, the characteristic scale $\alpha$ still exhibits a pronounced bimodal structure. This indicates that the S2 star only provides a tomographic scan of the environmental matter halo, without being able to determine its characteristic scale.
Resolving this degeneracy requires combining multiple stars with orbital radii both inside and outside that of the S2 star, such as the inner stars S4714, S301, S4716, etc.\cite{Pei_ker_2020,Dayem:2026ktt,Peißker_2022}, and the outer stars S1, S38, S55, etc.\cite{Gillessen:2017jxc,Meyer:2012hn}. Leveraging the different radial sensitivity regions probed by different stellar orbits, a multi-star joint analysis is expected to progressively reconstruct the radial distribution of the environmental matter and determine its true characteristic scale. In addition, weak-field tests within the Solar System \cite{Will:2014kxa,Fomalont:2009zg} can also provide complementary avenues for constraining environmental matter. 
With the improved observational precision of the ngEHT \cite{Johnson:2023ynn}, GRAVITY+ \cite{GRAVITY11}, and extremely large telescopes \cite{ELT11}, multi-probe and multi-star joint analyses can further tighten the upper limit on the environmental compactness $C$ and may identify the radial distribution profile of the environmental matter. When observational precision becomes sufficient to resolve black hole spin effects, the model discussed in this paper will need to be extended from a static spherically symmetric spacetime to an axisymmetric one.

\section*{acknowledgements}
This work was supported by the National Natural Science Foundation of China (Grant Nos. 12365008, 12265007), Guizhou Provincial Basic Research Program (Natural Science) (Grant No. QianKeHeJiChu[2024]Young166), the Guizhou Provincial Basic Research Program (Natural Science) (Grant Nos. QianKeHeJiChu-ZK[2024]YiBan027, QianKeHeJiChuMS[2025]680), the Guizhou Provincial Major Scientific and Technological Program XKBF (2025) 010 (Hosted by Professor Xu Ning), the Guizhou Provincial Major Scientific and Technological Program XKGF (2025) 009 (Hosted by Professor Xiang Guoyong), and the Guizhou Provincial Major Scientific and Technological Program (moderated by Teacher Fan Lulu).

\section*{Data availability}
All data involved in this study are openly available in published literature. Specifically, the black hole shadow data are taken from Ref. \cite{EventHorizonTelescope:2022wkp}; the orbital data of the S2 star are sourced from Do et al. \cite{Do:2019txf}, Gillessen et al. \cite{Gillessen:2017jxc}, and the GRAVITY Collaboration \cite{GRAVITY:2020gka}.

\appendix

\section{Geodesics} \label{a1}
This appendix presents the derivation of the geodesic equations in the equatorial plane for a static spherically symmetric spacetime. For the metric \eqref{eq:metric}, the Lagrangian can be written as
\begin{equation}
	\mathcal{L} = \frac{1}{2} g_{\mu\nu} \dot{x}^\mu \dot{x}^\nu,
	\label{eql}
\end{equation}
where the dot denotes the derivative with respect to the affine parameter $\lambda$, i.e., $\dot{x}^{\mu}\equiv dx^{\mu}/d\lambda$. For timelike geodesics, $\lambda$ can be taken as the proper time $\tau$ of the particle; for null geodesics, $\lambda$ is the affine parameter. The corresponding conjugate momenta are
\begin{equation}
	p_\mu\equiv \frac{\partial \mathcal{L}}{\partial \dot{x}^{\mu}}=g_{\mu\nu}\dot{x}^{\nu}.
\end{equation}

Since the background spacetime is static and spherically symmetric, there exist two corresponding Killing vectors: $\xi^\mu_{(t)}=(\partial_t)^\mu$ and $\xi^\mu_{(\phi)}=(\partial_\phi)^\mu$. Therefore, the conserved quantities along geodesics can be defined as
\begin{equation}
	\mathcal{E}\equiv -p_{\mu}\xi^\mu_{(t)}=-p_t,\qquad \ell\equiv p_{\mu}\xi^\mu_{(\phi)}=
	p_\phi ,
	\label{eq:EL_def}
\end{equation}
For the background geometry given by Eq. \eqref{eq:metric}, they take the form
\begin{equation}
	\mathcal{E}=A(r)\dot{t},
	\qquad
	\ell=r^2\sin^2\theta\,\dot{\phi}.
	\label{eq:EL}
\end{equation}
Here, $\mathcal{E}$ and $\ell$ can be interpreted as the energy per unit mass and the angular momentum per unit mass for massive particles; for photons, they are conserved quantities defined along the affine parameter.

Due to the spherical symmetry of the spacetime geometry, geodesic orbits can, without loss of generality, be taken to lie in the equatorial plane, i.e., $\theta=\frac{\pi}{2}$, $\dot{\theta}=0$. In this plane, Eq. \eqref{eq:EL} reduces to
\begin{equation}
	\dot{t}=\frac{\mathcal{E}}{A(r)},\qquad \dot{\phi}=\frac{\ell}{r^2}.
	\label{eq:EL1}
\end{equation}
Geodesics satisfy the normalization condition
\begin{equation}
	g_{\mu\nu}\dot{x}^{\mu}\dot{x}^{\nu}=-\kappa ,
	\label{eq:geo_constraint}
\end{equation}
where
\begin{equation}
	\kappa=
	\begin{cases}
		1, & \text{timelike geodesic}\\
		0, & \text{null geodesic}
	\end{cases}.
\end{equation}
Combining Eq. \eqref{eq:metric}, Eq. \eqref{eq:EL1}, and Eq. \eqref{eq:geo_constraint} in the equatorial plane yields the equation of motion for a test particle in the equatorial plane:
\begin{equation}
	\dot{r}^2=B(r)\left(  \frac{\mathcal{E}^2}{A(r)}-\frac{\ell^2}{r^2}-\kappa \right)
	\label{eq:radial}
\end{equation}

For photons, the motion is governed by null geodesics, i.e., $\kappa=0$. In this case, the radial equation \eqref{eq:radial} reduces to
\begin{equation}
	\dot{r}^{\,2}=B(r)\left[\frac{\mathcal{E}^2}{A(r)}-\frac{\ell^2}{r^2}\right].
	\label{eq:null_radial}
\end{equation}
Introducing the impact parameter $b\equiv \ell/\mathcal{E}$, the above equation can be further rearranged as
\begin{equation}
	\dot{r}^{\,2}=B(r)\mathcal{E}^2\left[\frac{1}{A(r)}-\frac{b^2}{r^2}\right].
\label{eq:null_radial1}
\end{equation}
The photon sphere corresponds to a circular orbit with constant radius $r_{\rm ph}$, satisfying the conditions
\begin{equation}
	\dot{r}^2\bigg|_{r=r_{ph}}=0,\; \frac{d \left(\dot{r}^{\,2}\right)}{dr}\bigg|_{r=r_{ph}}=0,\; \frac{d^2 \left(\dot{r}^{\,2}\right)}{dr^2}\bigg|_{r=r_{ph}} > 0.
	\label{eq:photon}
\end{equation}
From the first condition of Eq. \eqref{eq:photon}, we obtain
\begin{equation}
	b_{\rm ph}^2=\frac{r_{\rm ph}^2}{A(r_{\rm ph})}.
	\label{eq:bph}
\end{equation}
The second condition yields the equation for the photon sphere radius:
\begin{equation}
	r_{\rm ph}A'(r_{\rm ph})-2A(r_{\rm ph})=0.
	\label{eq:photon_sphere}
\end{equation}
In this paper, we mainly employ Eqs. \eqref{eq:bph} and \eqref{eq:photon_sphere} to determine the critical scale of the black hole shadow and calculate its theoretical diameter, which is then incorporated as a model prediction into the MCMC analysis.

\begin{figure*}[htbp]
	\centering
	\includegraphics[width=1 \textwidth]{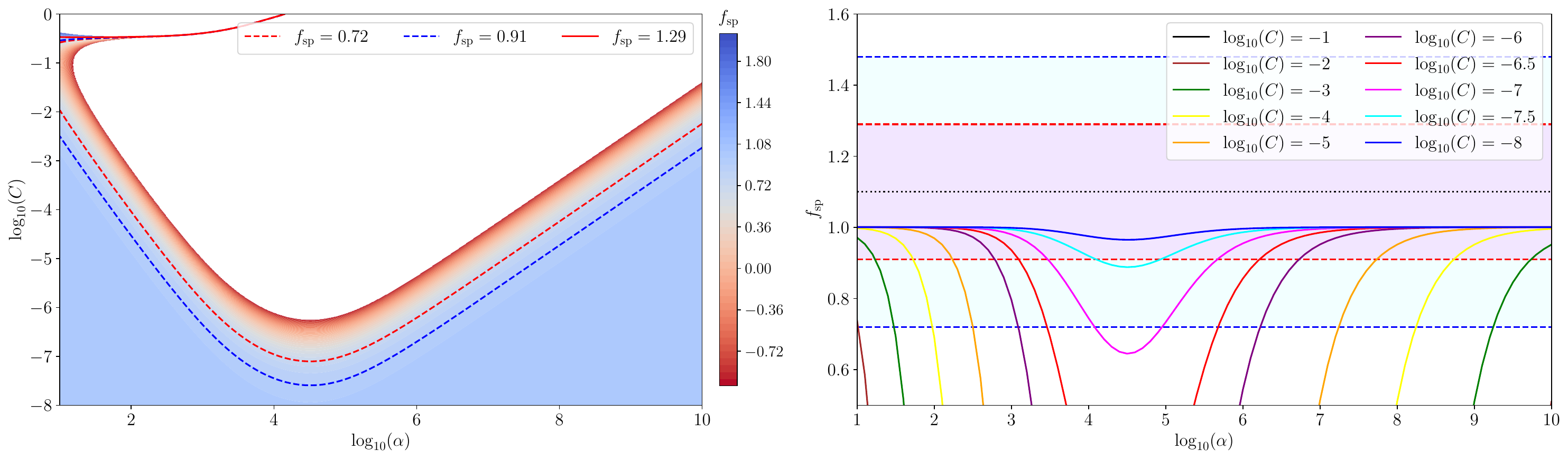}
	\caption{Constraints on parameters $C$ and $\alpha$ from observational data. Dashed lines correspond to the $1\sigma$ and $2\sigma$ confidence regions.}
	\label{constraint1}
\end{figure*}

For massive test particles, the motion is governed by timelike geodesics, i.e., $\kappa=1$. The radial equation of motion is
\begin{equation}
	\dot{r}^{2}=	B(r)\left[\frac{\mathcal{E}^2}{A(r)}-\frac{\ell^2}{r^2}-1\right].
	\label{eq:timelike_radial}
\end{equation}
Using $u = 1/r$ together with Eq. \eqref{eq:EL1}, Eq. \eqref{eq:timelike_radial} can be further transformed into a second-order orbital equation for $u(\phi)$:
\begin{align}
\frac{d^2 u}{d\phi^2} &= 
\frac{\mathcal{E}^2 \left( B_u A'_u - A_u B'_u \right) + A_u^2 \left[ \left( 1 + \ell^2 u^2 \right) B'_u - 2\ell^2 u^3 B_u \right]}
	{2\ell^2 u^2 A_u^2} \notag \\
&\equiv  F(u),
\end{align}
where
\begin{align}
&A_u = A(1/u),\quad B_u = B(1/u),\notag \\
&A'_u = \left.\dfrac{dA}{dr}\right|_{r=1/u},\quad B'_u = \left.\dfrac{dB}{dr}\right|_{r=1/u}.
\end{align}
Furthermore, the evolution of the azimuthal angle with time can be derived from Eq. \eqref{eq:EL1} as
\begin{equation}
	\frac{d\phi}{d t}=\frac{A_u \ell u^2  }{\mathcal{E}}.
\end{equation}

For the orbital precession of the S2 star, to relate the observed orbital parameters to the conserved quantities of timelike geodesics, we adopt a Keplerian-like radial parametrization, which is given by

\begin{equation}
	r=\frac{a(1-e^2)}{1+e \cos\psi}.
	\label{orbital}
\end{equation}
Here, $a$ is the semimajor axis, $e$ is the eccentricity, and $\psi$ is the radial phase angle. This parametrization is only used to determine the radial turning points and does not assume that the orbit is a closed Keplerian ellipse. The radii of the pericenter and apocenter are
\begin{equation}
	r_p=a(1-e),\quad r_a=a(1+e).
\end{equation}
At the two turning points, we have
\begin{equation}
	\dot{r}(r_p) = \dot{r}(r_a) = 0.
\end{equation}
Substituting these conditions into the radial equation of motion \eqref{eq:timelike_radial} yields
\begin{equation}
	\mathcal{E}^2= \frac{[r_a^2 - r_p^2] A(r_a) A(r_p)}{  r_a^2 A(r_p)-r_p^2 A(r_a)} ,
\end{equation}
\begin{equation}
	\ell^2= \frac{r_a^2 \left[ -r_p^2 A(r_a) + r_p^2 A(r_p) \right]}{r_p^2 A(r_a) - r_a^2 A(r_p)}.
\end{equation}
Therefore, given $a$, $e$, and the background metric $A(r)$, the first integrals $\mathcal{E}$ and $\ell$ of the timelike geodesic can be determined from the pericenter and apocenter conditions, and serve as inputs for the orbital integration and precession calculation of the S2 star.

\section{Precession Constraints}\label{B}

The orbital precession angle of the S2 star can be obtained entirely from the integration of timelike geodesics. From Eqs. \eqref{eq:EL1} and \eqref{eq:timelike_radial}, we have
\begin{equation}
	\frac{d \phi}{dr}=\frac{\ell}{r^2\sqrt{B(r)\left[\frac{\mathcal{E}^2}{A(r)}-\frac{\ell^2}{r^2}-1\right]}}.
\end{equation}
The pericenter precession angle over one complete radial period can be expressed as
\begin{equation}
	\Delta\phi = 2 \int_{r_p}^{r_a} \frac{\ell \, dr}{r^2 \sqrt{B(r)\left[ \frac{\mathcal{E}^2}{A(r)} - \frac{\ell^2}{r^2} - 1 \right]}} - 2\pi.
	\label{eq:phi}
\end{equation}
Differentiating the parametric equation \eqref{orbital} gives
\begin{equation}
	\frac{dr}{d\psi} = \frac{a(1-e^2)e\sin\psi}{(1+e\cos\psi)^2}.
\end{equation}
Substituting this into Eq. \eqref{eq:phi} yields
\begin{equation}
	\Delta\phi = 2 \int_{0}^{\pi} \frac{\ell e \sin\psi \, d\psi}
	{a(1-e^2)\sqrt{B_{\psi}\left[\dfrac{\mathcal{E}^2}{A_{\psi}} - \dfrac{\ell^2(1+e\cos\psi)^2}{a^2(1-e^2)^2} - 1\right]}} - 2\pi,
	\label{eq:phi1}
\end{equation}
where $A_{\psi} = A[r(\psi)]$ and $B_{\psi} = B[r(\psi)]$.

For the Hernquist-type environmental black hole spacetime discussed in this paper, the precession angle can be directly obtained by numerically integrating Eq.  \eqref{eq:phi1}. The corresponding deviation from the Schwarzschild black hole can be expressed as
\begin{equation}
	f_{sp}=\frac{\Delta \phi}{\Delta \phi_{\text{Sch}}},
\end{equation}
where $\Delta \phi_{\text{Sch}}$ is the precession angle of the Schwarzschild black hole. At the first post-Newtonian order, it is given by
\begin{equation}
	\Delta \phi_{GR}=\frac{6\pi M_{\rm BH}}{a(1-e^2)}.
\end{equation}
The GRAVITY Collaboration measured the relativistic precession parameter of the S2 star  as \cite{GRAVITY:2020gka}

\begin{equation}
	f_{sp}=\frac{\Delta \phi}{\Delta \phi_{GR}}=1.1\pm0.19.
\end{equation}

Using the observational data, the parameters $C$ and $\alpha$ can be preliminarily constrained. Here we use the latest reported data of the S2 star orbiting Sgr A* from the GRAVITY collaboration \cite{GRAVITY:2021xju}: $M_{\rm BH}=4.297\times 10^6 M_\odot$, $D=8277$ pc, $a=124.95$ mas, and $e=0.88441$. 
As shown in Fig. \ref{constraint1}, in the regions of small and large $\alpha$, a wide range of $C$ is compatible with the observationally allowed region; near $\log_{10}\alpha\sim4$, the precession correction is more sensitive to $C$, and only small $C$ remains allowed. As a result, the allowed region is compressed around intermediate scales, but is not completely broken.


\bibliography{ref}
\bibliographystyle{apsrev4-1}

\end{document}